\documentclass[10pt]{article}

\usepackage{amsmath}
\usepackage{latexsym}
\usepackage{amsfonts}
\usepackage{amssymb}

\usepackage{graphicx}

\usepackage{bbm,dsfont}

\title{QUANTUM MEASUREMENTS CONSTRAINED BY SYMMETRIES}

\author{P. BUSCH\thanks{E-mail: {\tt paul.busch@york.ac.uk}}
\\
{\small Department of Mathematics}\\
{\small University of York, York YO10 5DD, UK}\\
\phantom{XXX}\\
L. D. LOVERIDGE\thanks{E-mail: {\tt Leon@phas.ubc.ca}}\\
{\small Department of Physics and Astronomy}\\
{\small University of British Columbia,
Vancouver, BC V6T 1Z1 Canada}
}

\date{}

\begin{document}
\maketitle

\noindent
{\small {\em The XXIXth International Colloquium on Group-Theoretical Methods in Physics}, 
Chern Institute of Mathematics, August 20-26, 2012. {\em Nankai Series in Pure, Applied Mathematics and Theoretical Physics},
World Scientific, 2013.}

\begin{abstract}

We revisit the theorem of Wigner, Araki and Yanase (WAY) describing limitations to  repeatable quantum measurements
that arise from the presence of conservation laws. We will review a strengthening of this theorem by exhibiting and discussing
a condition that has hitherto not been identified as a relevant factor. We will also show that an extension of the theorem to continuous
variables such as position and momentum can be obtained if the degree of repeatability is suitably quantified. \\


\end{abstract}

\thispagestyle{empty}

\section{Introduction}\label{sec:intro}

Measurements are physical processes and as such are subject to the laws of physics. In particular,
the interaction between a measurement device or probe and an object system needed to establish
suitable correlations between the quantity being measured and a pointer observable are constrained
by fundamental symmetries, or equivalently the associated conservation laws. In quantum mechanics this has
been found to lead to restrictions of the measurability of certain physical quantities, as demonstrated
first by Wigner \cite{Wigner1952} and then in more generality by Araki and Yanase \cite{AY1960}.

Here we will review a strengthening and an an extension of these results in two respects. First
we show that the presence of an additive conservation law not only precludes {\em repeatable, sharp} 
measurements of a discrete quantity not commuting with the system
part of the conserved quantity, but in addition a further property of a measurement must be violated:
the so-called Yanase condition, which states that the pointer observable commutes with the conserved quantity (Sec. 3).
Second, a WAY-type limitation is shown to hold for
position measurements obeying momentum conservation, thus extending the WAY theorem to continuous
quantities (Sec. 4). We begin with a brief account of the formal theory of quantum measurement as far as required
for the formulation of these results.

\section{Quantum Measurements}

In the standard formulation of quantum mechanics, a physical system is associated with a Hilbert space 
$\mathcal{H}$ over $\mathbb C$; its states are represented as density operators $\rho$, with pure states 
as normalized vectors $\psi\in{\mathcal H}$. Observables are given as positive operator valued measures (POVMs) 
${\sf E}$ from a $\sigma$-algebra $\Sigma$ of subsets of the set of measurement outcomes to the set 
of effects (positive operators), $ X\mapsto {\sf E}(X)$. The probability that a measurement of 
${\sf E}$ in a state $\rho$ gives an outcome in the set $X$  then is  ${\sf p}_\rho^{{\sf E}}(X)={\rm tr}\bigl[\rho \,{\sf E}(X)\bigr]$.

As an example, we refer to an {\em unsharp}, or {\em smeared} position observable, defined as follows:
\begin{align}\label{eqn:sp}
{\sf Q}^e(X)&=\int_{\mathbb R} e(x){{\sf E}}^Q(x+X)dx=\chi_X*e(Q),  \\
\langle\varphi|{\sf Q}^e(X)\varphi\rangle&={\sf p}_\varphi^{{\sf Q}^e}(X)=
\int dx \left\vert \varphi
(x)\right\vert ^{2} \chi _{X} * e(x),
\end{align}
where ${\sf E}^Q$ denotes the spectral measure of the position operator $Q$, $\chi_X$ a characteristic (set indicator)
function, $*$ denotes convolution, and  $e$ is a distribution function representing the measurement inaccuracy. 
Such POVMs occur in a variety of models of approximate position measurements.

The time evolution of an isolated system is given by a unitary group $U_t$ via
$\rho_t=U_t^{}\rho_0U_t^*$, $U_t^{}=\exp(-iHt/\hbar)$, or the associated Schr\"odinger equation, 
$i\hbar\, d\psi_t/dt=H\psi_t$.

A measurement is modelled as an interaction between the object system (represented by ${\mathcal H}$) and an
apparatus or probe system (represented by $\mathcal{K}$), such that the time evolution $U$ acts 
on the tensor product ${\mathcal H}\otimes \mathcal{K}$.
Thus, a measurement ${\mathcal{M}}$ of an observable ${\sf E}$ is summarized as a quintuple 
$\langle \mathcal{K},U,\phi ,{ Z},h\rangle $, where $\phi$ is the initial probe state, ${ Z}$ the pointer observable
(normally modelled as a selfadjoint operator with associated spectral measure $\mathsf{E}^{Z}$), and 
$h$ a pointer scaling function mapping the pointer values to
the values of the measured observable. The latter is then determined by the {\em probability reproducibility condition},
according to which the coupling $U$ sending the initially uncorrelated system-probe state to an entangled state,
$$
\Psi_i\equiv\varphi\otimes\phi\to\Psi _{f }\equiv U(\varphi \otimes \phi )\in {\mathcal H}\otimes\mathcal{K},
$$
serves as a measurement of the observable ${\sf E}$ if the pointer probabilities reproduce the probabilities of ${\sf E}$:
\begin{equation}
\left\langle \Psi _{f }|{\mathbbm{1}}\otimes {\sf E}^{Z}(h^{-1}(X))\Psi _{f}
\right\rangle \equiv \left\langle \varphi |{\sf E}(X)\varphi \right\rangle.
\label{PRC}
\end{equation}
A measurement is called {\em repeatable} if
\begin{equation}\label{rep}
\left\langle \Psi _{f} |{\sf E}(X)\otimes {\sf E}^{Z}(h^{-1}(X))\Psi _{f}
\right\rangle =\left\langle \varphi |{\sf E}(X)\varphi \right\rangle.
\end{equation} 
The system state conditional on an outcome (in) $X$ is defined via
\begin{equation*}
\rho_X={\rm tr}_{\mathcal K}\left[{\mathbbm{1}}\otimes {\sf E}^{Z}\left(h^{-1}(X)\right)|\Psi _{f}\rangle\langle \Psi _{f}|\right];\quad
\hat\rho_X=\frac{\rho_X}{{\rm tr}[\rho_X]}.
\end{equation*}
Then the probability reproducibility and repeatability conditions are compactly given as 
 ${\rm tr}[\rho_X]=\left\langle \varphi |{\sf E}(X)\varphi \right\rangle$
and 
${\rm tr}[\hat\rho_X\,{\sf E}(X)]=1$, 
respectively.

It is a historical curiosity that until rather recently the term measurement was understood by many as including the property
of repeatability. Only very few authors, such as Pauli, Landau and Peierls in the 1930s and Wigner, allowed for the 
possibility of measurements that are not necessarily repeatable.

\section{The Theorem of Wigner, Araki and Yanase}

In 1952, Wigner \cite{Wigner1952} considered a spin-$\frac12$ system and showed that a repeatable measurement of the spin component
$s_x$ violates conservation of (say) the $z$-component of the total angular momentum shared between system and probe, $s_z+J_z$.
He also showed that if the probe's angular momentum $J_z$ is allowed to have a large spread in it's initial state $\phi$, 
then an arbitrarily accurate and repeatable 
measurement of $s_x$ becomes possible. Indeed, with a measurement coupling:
\begin{align*}
\varphi _{+}\otimes \phi &\longrightarrow \varphi _{+}\otimes \phi
_{+}+\varphi _{-}\otimes \eta , \quad
\varphi _{-}\otimes \phi \longrightarrow \varphi _{-}\otimes \phi
_{-}-\varphi _{+}\otimes \eta 
\end{align*}
(and $\left\langle \eta |\phi _{\pm }\right\rangle = \left\langle \phi_+ |\phi _{- }\right\rangle = 0$),
the measured observable has {\em three} effects: 
\[
E_{\pm }=(1- \left\Vert \eta \right\Vert ^2) P[\varphi_{\pm}], \quad E_{?}=\left\Vert\eta \right\Vert ^2{\mathbbm{1}}.
\]
Here $\varphi_\pm$ are the eigenstates of the measured spin component $s_x$ and $\phi_\pm,\eta$ are pointer eigenstates. The third
effect is trivial: it provides no information as its expectation value does not depend on the object state. The associated outcome is
`uncertain'.

Wigner noted that the conservation law can be satisfied in the above dynamics while $\left\Vert\eta \right\Vert ^2=1/(2n-1)$, or even 
$\left\Vert\eta \right\Vert ^2\sim \frac1{n^2}$, if $\phi =\sum\limits_{\nu =1}^{n}\phi _{\nu}$ 
(where $\phi_\nu$ are $J_z$-eigenvectors). For large $n$, one obtains $E_{\pm }\simeq P[\varphi_{\pm}]$, $E_{?}\simeq {\mathbbm{O}}$.

Araki and Yanase \cite{AY1960} generalized Wigner's result to the effect that any observable that allows a repeatable measurement in the presence
of an additive conserved quantity must commute with the object part of that conserved quantity. This statement is subject to the assumption that
the measured observable is discrete (thus allowing repeatable measurements) and at least the object part of the conserved quantity is bounded.
Yanase \cite{Yanase1961} constructed a model in which, in analogy to Wigner's observation, arbitrary accuracy and good repeatability are
obtained if large variances in the probe part of the conserved quantity are allowed.  This followed even under an additional assumption, now
known as the {\em Yanase condition}: that the pointer observable commutes with the probe part of the conserved quantity. This can be seen
as a pre-emptive assumption that ensures that the WAY measurement obstruction does not reappear in the measurement of the pointer.

It was shown recently that besides repeatability, the Yanase condition cannot be satisfied in a sharp measurement of an observable that
does not commute with an additive conserved quantity.\cite{LB2011}  This is the strengthened version of the WAY theorem.

\section{Position measurement vs. momentum conservation}

It has remained an open question for a long time whether the WAY theorem extends to object observables with continuous spectra. In that
case repeatability cannot be obtained since according to a fundamental theorem due to Ozawa \cite{Ozawa1984}, the existence of repeatable
measurements requires the observable to be discrete. However, it is known that approximate repeatability is still available in the general case.

One can thus investigate whether momentum conservation entails an obstruction to approximately repeatable position measurements. This was
denied by Ozawa\cite{Ozawa1991} in 1991, who presented a model of a momentum-conserving position measurement. Closer scrutiny of this
model (which satisfies the Yanase condition) has revealed that good accuracy and repeatability come at the expense of a large variance of the probe's momentum.
Indeed, the measured observable is of the form \eqref{eqn:sp}, and it can be seen that the function $e$
can be made narrow only by increasing the momentum spread in a state corresponding to the preparation of the apparatus.
This can be seen
as a quantitative version of the WAY theorem, and it has been shown by the present authors\cite{BL2011} that there is a general trade-off relation between 
the quality of repeatability and the variance of the probe momentum under the satisfaction of the Yanase condition. This will be briefly reviewed here.

A common way of quantifying the difference between the observable one aims to measure approximately, represented by selfadjoint operator $M$,
and the actually measured observable is by means of the {\em noise operator}, $N=Z_f-M$,  where $Z_f =U^*{\mathbbm{1}}\otimes ZU$ and $Z$ is the 
selfadjoint pointer observable. The {\em noise}
$\epsilon(\varphi)$ is then given by 
$\epsilon (\varphi )^{2}:=\left\langle \varphi \otimes \phi |N^{2}\varphi \otimes \phi \right\rangle \equiv \langle N^{2}\rangle$.  
Noting that $\langle N^{2}\rangle\geq (\Delta N)^2$ and using the Cauchy-Schwarz inequality, Ozawa\cite{Ozawa2002} obtained the following inequality:
\begin{equation}\label{Ozawa_error}
\epsilon (\varphi )^{2}\geq 
\frac{1}{4}\frac{\left\vert \left\langle \left[Z_f-M,L_{1}+L_{2}\right] \right\rangle \right\vert ^{2}}{(\Delta _{\varphi }L_{1})^{2}+(\Delta _{\phi }L_{2})^{2}}.  
\end{equation}
Here $L_1+L_2$ is the additive conserved quantity. Stipulating the Yanase condition, the numerator becomes 
$\left\vert \left\langle \left[M,L_{1}\right] \right\rangle \right\vert ^{2}$. We can then estimate the {\em global error},
$\epsilon:=\sup_\varphi\epsilon(\varphi)$ and obtain:
\begin{equation}\label{Ozawa_error2}
\epsilon^{2}\geq \sup_{\varphi}\frac{1}{4}\frac{\left\vert \left\langle \left[M,L_{1}\right] \right\rangle \right\vert ^{2}}{(\Delta _{\varphi }L_{1})^{2}+(\Delta _{\phi }L_{2})^{2}}.  
\end{equation}
This can be applied to position measurements obeying momentum conservation, where $M=Q$ and $L_1=P$, $L_2=P_{\mathcal A}$, giving
\begin{equation}\label{LB-tradeoff}
\epsilon^{2}\geq \frac{\hbar^2}{4(\Delta_\phi P_{\mathcal A})^2}.
\end{equation}
Similarly one can define a measure of the degree of repeatability by replacing  $M$ in the definition of the noise operator by the Heisenberg-evolved
$M_f=U^*M\otimes{\mathbbm{1}} U$:
 $\mu (\varphi )^{2}:=\langle \varphi \otimes \phi |(M_f-Z_f)^{2}\varphi \otimes \phi \rangle$.
 For $\mu :=\sup\mu (\varphi )$ one obtains: 
\begin{equation}\label{rep-tradeoff}
\mu ^{2}\geq\sup_{\varphi } \frac{1}{4}\frac{\left\vert \left\langle \left[ M(\tau )-Z(\tau
),L_{1}+L_{2}\right] \right\rangle \right\vert ^{2}}{(\Delta_{\varphi }
L_{1})^{2}+(\Delta _{\phi }L_{2})^{2}} . 
\end{equation}
With the
Yanase condition the numerator becomes $\left\vert \left\langle \left[ M_f,L_{1}+L_{2}\right] \right\rangle \right\vert ^{2}$. In the case of
position measurements the above inequality then gives
\begin{equation}
\mu^{2}\geq \frac{\hbar^2}{4(\Delta _{\phi }P_{\mathcal A})^{2}} .
\end{equation}

A first conclusion to be drawn is that under the Yanase condition, momentum conserving position measurements can only be
performed with good accuracy and repeatability if the probe system is `large', in the sense that its momentum variance must be
large. This limitation is confirmed with Ozawa's position measurement model.\cite{BL2011}

The case of position and momentum allows for a particular way of violating the Yanase condition, so that the numerators of
inequalities (\ref{Ozawa_error}) and (\ref{rep-tradeoff}) become zero: this happens when the pointer is chosen to be the position 
$Q_{\mathcal A}$. In this case there is no size constraint on the probe that needs to be fulfilled in order that $\epsilon$ and $\mu$ 
can be made small, as can be confirmed by another position measurement model.\cite{BL2011} However, as noted above, the 
problem of measuring position in view of momentum conservation has now been shifted to the probe system. Thus a second conclusion
is that information transfer about the position of an object to a probe system is perfectly possible if one relinquishes the Yanase condition,
but whether this can be used for measurements depends on the measurability of the probe position under the constraint of momentum
conservation.

Finally we note that if instead of position one seeks to measure the relative position between object and probe $M=Q-Q_{\mathcal A}$, 
then the numerators in (\ref{Ozawa_error}) and (\ref{rep-tradeoff}) vanish if the Yanase condition holds since the relative position
commutes with the total momentum. Closer inspection of measurement models shows that it is not primarily the large variance of the
probe momentum that one needs to seek to achieve to obtain good measurement of the object position alone. It is rather the fact that
the relative position  effectively becomes indistinguishable from the object position if the probe position is prepared  with low inaccuracy;
but this entails large variance of the probe momentum by virtue of the uncertainty relation.\cite{Loveridge2012}

\section*{Acknowledgments}
This work was supported by a doctoral training grant from EPSRC (UK).


\begin{thebibliography}{9}

\bibitem{Wigner1952} E.~Wigner,  
\emph{Z.~Phys.} \textbf{133}, 101 (1952). English translation: arXiv:1012.4372.


\bibitem{AY1960} H.~Araki, M.M.~Yanase,  
\emph{Phys.~Rev.} \textbf{120}, 622 (1960).

\bibitem{Yanase1961} M.M.~Yanase, \emph{Phys.~Rev.}~\textbf{123}, 666 (1961).

\bibitem{LB2011} L.D. Loveridge, P. Busch, {\em Eur.~Phys.~J.~D} {\bf 62}, 297 (2011).

\bibitem{Ozawa1984} M.~Ozawa,  \emph{J.~Math.~Phys.}~\textbf{25}, 79 (1984).


\bibitem{Ozawa1991} M.~Ozawa,  \emph{Phys.~Rev.~Lett.}~\textbf{67}, 1956 (1991).

\bibitem{BL2011} P. Busch, L.D. Loveridge, {\em Phys.~Rev.~Lett.} {\bf 106}, 110406 (2011).

\bibitem{Ozawa2002} M. Ozawa, \emph{Phys.~Rev.~Lett.}~\textbf{88}, 050402 (2002).

\bibitem{Loveridge2012} L.D. Loveridge, {\em Quantum Measurements in the Presence of Symmetry}, Doctoral Thesis,
University of York, 2012 (http://etheses.whiterose.ac.uk/2670).


\end{thebibliography}
\end{document}